\documentclass[12pt]{article}
\font \greekb=cmmib9 scaled \magstep1
\newcommand{\Sigmab}{\mbox{\greekb \char 6}}
\begin{document}

\begin{center}
{\Large \bf Spin rotation as an element of polarization experiments on elastic
electron-proton scattering}

\vspace{0.5 cm}

\begin{small}
\renewcommand{\thefootnote}{*}
L.M.Slad\footnote{slad@theory.sinp.msu.ru} \\

\vspace{0.3 cm}

{\it Skobeltsyn Institute of Nuclear Physics,
Lomonosov Moscow State University, Moscow 119991, Russian Federation}
\end{small}
\end{center}

\vspace{0.5 cm}

\begin{footnotesize}
The validity of some rules in the classical theory of spin, which are followed 
by the Bargmann--Michel--Telegdi formula for a relativistic particle spin
rotation in a constant homogeneous magnetic field, is analyzed. In the 
framework of the quantum theory, we give examples where these rules are 
violated.  Consequences for polarization experiments are discussed.
\end{footnotesize}

\vspace{0.5 cm}

\begin{small}

\begin{center}
{\large \bf 1. Introduction}
\end{center}

In the 2000s, a series of unique experiments on elastic electron-proton
scattering have been carried out at Jefferson Laboratory [1]--[7]. 
A substantial discrepancy in the values of the ratio of electric $G_{E}$ and 
magnetic $G_{M}$ formfactors of the proton is found for two different 
techniques of extracting them from experimental measurements. In one of the 
approaches, the values of the quantity $R = \mu_{p}G_{E}/G_{M}$ (with $\mu_{p}$
being the magnetic moment) is extracted from polarization experiments 
[1]--[5] using the Akhiezer--Rekalo formula [8]. They decrease almost linearly from unity to approximately 0.3 when the squared transferred momentum $Q^{2}$ increases from 0 to 5.6 (GeV/c)$^{2}$. In the other approach, the values of $R$ are obtained from the Rosenbluth formula [9] at processing of the high-precision unpolarized experiments [6], [7]. These values are close to unity in the same interval of $Q^{2}$.

The results of old numerous calculations of radiative corrections to the 
Rosenbluth formula, which are summarised in Ref. [10], have a rather 
small influence on the values of $R$ [11]. A new analysis 
[12]--[16] of the possible size of two-photon exchange contributions to the Rosenbluth and Akhiezer--Rekalo formulae shows, according to the dominant opinion, that the discrepancy between the mentioned values of 
$R$ can be reduced to some extent, but not eliminated at all.

In view of the above, we find it necessary to carefully reconsider the
theoretical basis used in deriving the final results in the experimental works 
[1]--[7]. The essential ingredients of this basis are (i) assuming 
the proton to be a Dirac particle, that realizes in the general form of the 
nucleon electromagnetic current [17] and, further on, in the Rosenbluth 
and Akhiezer--Rekalo formulae; (ii) using of the Bargmann--Michel--Telegdi 
(BMT) formula [18] to describe the relativistic proton spin rotation in a 
constant homogeneous magnetic field; and (iii) modelling the azimuthal 
asymmetry in the angular distribution of the protons after their secondary 
scattering on the carbon target as a result of spin-orbital interaction 
[19].

A strict analysis of some consequences of changing the general form of the 
nucleon electromagnetic current, caused by refusal of the description of the
proton and the neutron by a Dirac spinor, is presented in Ref. [20]. It is established that this change in itself cannot cause the appearance of the 
discussed contradictions in the values of $R$. Namely, the obtained formulae 
for the angular distribution of electrons elastically scattered on the protons, 
and, also for the ratio of the transversal and longitudinal components of the 
recoil proton polarization vector come accordingly to the Rosenbluth and 
Akhiezer--Rekalo formulae, irrespective of the representation of the proper 
Lorentz group $L^{\uparrow}_{+}$ attributed to the proton as a particle with 
the rest spin 1/2.

Let us proceed now with discussing the quantum theoretical validity of  
preconditions of the BMT formula.

\begin{center}
{\large \bf 2. Some rules of the classical theory of spin and their violation in the quantum theory}
\end{center}

Bargmann, Michel, and Telegdi [18] have obtained their formula assuming 
that, in the framework of classical theory, it is admissible to describe the 
spin with an axial 4-vector $s^{\alpha}$ (it was introduced by Tamm [21]), whose time component in the particle 
rest frame is zero
\begin{equation}
s^{\alpha}u_{\alpha} = 0, \label{1}
\end{equation}
where $u^{\mu}$ is the particle 4-velocity. On the other hand, the equation for
evolution of the spin at motion of a particle in an electromagnetic field has 
been originally proposed by Frenkel [22] and was expressed in terms of the properly introduced antisymmetric tensor of spin $s^{\mu\nu}$. According to 
Frenkel's rule, the tensor of spin has only space-space components in the 
particle rest frame, i.e.
\begin{equation}
s^{\mu\nu}u_{\nu} = 0. \label{2}
\end{equation}

In the classical theory, as we will see later, the rule (\ref{2}) is the 
necessary condition, first, for deriving the spin rotation formula based on the
spin tensor $s^{\mu\nu}$, and, second, for the consistent description of spin 
with 4-vector $s^{\alpha}$, including the equality (\ref{1}) which plays the 
key role in deriving the BMT formula in Ref. [18]. Since the spin is a 
group-theoretical concept, all our subsequent discussions and conclusions about
the consistency of this double description of the spin in the classical theory,
as well as the possibility to fulfill Eq. (\ref{2}) will be based on the 
group-theory definitions and exact solutions in the quantum field theory. Note 
that substantiating the BMT equation for a particle with rest spin 1/2 and 
arbitrary anomalous magnetic moment on the basis of a quantum description of 
that particle with Dirac equation have been attempted in Refs. [23], 
[24]. This substantiation cannot be said convincing, because of many 
assumptions introduced in the mathematical calculations.

As it is well known, in the nonrelativistic quantum mechanics, the concept of
spin is a group-theoretical analogue of the orbital angular momentum. Namely,
both the spin and orbital momenta associate with antisymmetric tensor 
operators, which are identified with the generators of the rotation group, the latter being, however, realized in essentially different ways in 
the spaces of group representations nonequivalent in their structure.  The 
orbital momentum corresponds to the realization of such generators as operators
of the form $-i(x^{i}\partial/\partial x^{j}-x^{j}\partial/\partial x^{i})$ in 
the space of coordinate functions. The spin refers initially to a matrix 
realization of the generators $L^{ij}$ of the rotation group in the space of 
finite-component vectors. Then, for a given particle state, the spin is 
described with the mean values of these generators forming an antisymmetric 
tensor $s^{ij}$ of the rotation group or an equivalent three-dimensional vector
${\bf s}= \{ s^{23}, s^{31}, s^{12}\}$. A natural extension of the above 
definition of the spin to the relativistic quantum theory consists in its
association with a matrix realization of the generators $L^{\mu\nu}$ of the 
proper Lorentz group in the space of any representation we are interested in. 
(Note that, sometimes, even in relativistic quantum theory as, for example,
within the framework of describing the irreducible representations of the group
$L^{\uparrow}_{+}$, the term of ``spin'' is connected only with the rotation 
group [25], [26].) The mean values of these generators can be 
interpreted for a given particle state as components of the antisymmetric 
tensor of spin $s^{\mu\nu}$ in the classical theory. As far as the 4-vector 
velocity $u^{\mu}$ is concerned, it has to be regarded in the quantum theory as
the mean value of the 4-momentum operator $P^{\mu}$ divided by the particle 
mass. For an arbitrary particle state not possessing a definite value of 
4-momentum, the rest system is given by the quantum mechanical value 
${\bf u} = 0.$

The quantum description of the spin given above and the classical description
of the spin which follows from that is logically simple and consecutive. It is
suitable for any particles and for any their states, and so, it should be
considered as the primary or standard description. Prior to employing the axial 
4-vector $s^{\alpha}$ in the classical theory of spin, we should be convinced,
that it is consistent with the standard description for all possible states 
under consideration. Namely, we have to make sure that, in all cases, there is
one-to-one correspondence between the tensor $s^{\mu\nu}$ and the 4-vector 
$s^{\alpha}$. Otherwise, the same word of "spin" will mean different physical
entities, that can lead to those or other contradictions.

The axial 4-vector $s^{\alpha}$ results from the antisymmetric tensor 
$s^{\mu\nu}$ by invoking the 4-velocity $u^{\mu}$ as 
\begin{equation}
s^{\alpha} = \frac{1}{2} \varepsilon^{\mu\nu\rho\alpha}s_{\mu\nu}u_{\rho}.
\label{3}
\end{equation}
From here, Eq. (\ref{1}) follows in particular.

Consider now the relation (\ref{3}) as a system of four linear equations with 
given values of quantities $s^{\alpha}$ and $u^{\mu}$ and with unknown values 
of six components of the antisymmetric tensor $s^{\mu\nu}$. We will be assuming
that the quantities $s^{\alpha}$ satisfy to the condition (\ref{1}), otherwise,
the system (\ref{3}) would be controversial. Then, one of the four equations in
(\ref{3}) is linearly expressed through three other. Thus, a general solution 
of this system for the unknown $s^{\mu\nu}$ contains three arbitrary constants 
and can be written as 
\begin{equation}
s^{\mu\nu} = \varepsilon^{\mu\nu\alpha\rho}s_{\alpha}u_{\rho}+u^{\mu}a^{\nu}-
u^{\nu}a^{\mu}, \label{4}
\end{equation}
where the 4-vector $a^{\mu}$, being arbitrary and noncollinear to $u^{\mu}$ 
($a^{\mu} \neq \lambda u^{\mu}$), is determined by the three constants 
mentioned above. It is easy to see that the tensor $s^{\mu\nu}$ is uniquely 
determined by the 4-vector $s^{\alpha}$, i.e. that $a^{\mu} = 0$, if and only 
if the relation (\ref{2}) is fulfilled.

There is no physical reasoning for any concrete quantum description of the 
spin with the axial 4-vector $s^{\alpha}$. All the suggestions on the choice of
the operator associated with the 4-vector $s^{\alpha}$ were only motivated by
appealing to the transformation properties of that operator under the
orthochronous Lorentz group.

Let us dwell in brief on the interpretation [18] of the axial 4-vector of spin $s^{\alpha}$ in the classical theory as the mean value of operator
\begin{equation}
W^{\alpha} = \frac{1}{2m} \varepsilon^{\mu\nu\rho\alpha}L_{\mu\nu}P_{\rho},
\label{4a}
\end{equation}
which was used by Bargmann and Wigner [27] to classify irreducible representations of the Poincar\'{e} group. (We would like to draw attention to that in Ref. [27] the antisymmetric tensor operator in matrix realization rather than the operator $W^{\alpha}$ is put in correspondence to the spin.) For consequences of the relation (\ref{4a}) an essential role is played by the fact that the mean of a product of two operators is equal to the product of their mean values only when a state, on which averaging is carried out, is the proper state for one of these operators, and besides, maybe, in some exceptional cases. Therefore, from formula (\ref{4a}) for operators the relation (\ref{3}) for classical quantities follows up only for states with definite value of the 
4-momentum, and, maybe, for some special states. For all other states, the operator relation (\ref{4a}) does not lead to the $c$-numerical relation with the structure given by formula (\ref{3}), also as the operator relation 
$W^{\alpha}P_{\alpha} = 0$, received from (\ref{4a}), does not lead to the relation  (\ref{1}). We will demonstrate it later on two examples showing that the quantity $s^{0}$, as mean value of the operator $W^{0}$ (\ref{4a}), is not equal to zero in the quantum mechanical rest system. Let us note also that, considering the relation (\ref{4a}) as the equation concerning the unknown antisymmetric operator $L^{\mu\nu}$, we receive the solution having the structure of type (\ref{4}) and containing an arbitrary 4-vector operator.

Another available interpretation [23], [24] of the spin 4-vector 
$s^{\alpha}$ assigned to a Dirac particle as the mean value of the operator 
$(1/2) \gamma^{5} \gamma^{\alpha}$ is, most likely, only restricted to free 
particle states with definite 4-momentum, when it is consistent, by direct 
calculations, with the standard description of the spin. It is easy to be convinced that, for the written below states (\ref{5})--(\ref{7}) and 
(\ref{10})--(\ref{12}), this treatment of the spin results in breaking the rule (\ref{1}).

It is worth saying in addition, that specifying of matrix 4-vector operators in
the quantum theory for a broad set of representations of the proper Lorentz 
group is non-unique. Indeed, if such an operator couples more than two 
irreducible representations of the group $L^{\uparrow}_{+}$, then it contains 
arbitrary constants which cannot be eliminated by any normalization [25], [26].

Let us now turn to the problem of feasibility of the equality (\ref{2}). 

It has been demonstrated in [20] that, for a broad class of finite- or 
infinite-dimensional representations of the group $L^{\uparrow}_{+}$ which 
correspond to the wave vectors of a particle, the following statement is valid.
If the rest state of a free particle, described by the vector $\psi(x) = 
\exp(-imt) \psi(p_{0})$, possesses definite parity (it is true for all states 
obeying this or that relativistic invariant equation), then the mean value of 
generators $L^{0i}$ ($i=1,2,3$) of the proper Lorentz group is zero, i.e. the 
relation (\ref{2}) holds.

Consider now the state of a free Dirac particle, representing a wave packet in 
the momentum space
\begin{equation}
\psi = (2\pi )^{-3/2} N^{1/2} \int \exp(-iEt+i{\bf p}{\bf r})
\left[ c_{+1}({\bf p}) \psi_{+1}({\bf p})+
c_{-1}({\bf p}) \psi_{-1} ({\bf p}) \right] d{\bf p}, \label{5}
\end{equation}
where $N$ is the normalization coefficient, $c_{\pm 1}({\bf p})$ are some 
arbitrary functions of momentum, and $\psi_{\pm 1}({\bf p})$ are the Dirac 
spinors, normalized to unity, which have in the spherical coordinate system of 
the momentum space the following form
\begin{eqnarray}
\psi_{+1}({\bf p}) = \frac{1}{\sqrt{2m}} \left(
\begin{array}{c}
\sqrt{E+m} \cos (\theta/2) \exp(-i\phi/2)  \\
\sqrt{E+m} \sin (\theta/2) \exp(i\phi/2)    \\
\sqrt{E-m} \cos (\theta/2) \exp(-i\phi/2)   \\
\sqrt{E-m} \sin (\theta/2) \exp(i\phi/2)   
\end{array} 
\right) , \label{6}
\\
\psi_{-1}({\bf p}) = \frac{1}{\sqrt{2m}} \left(
\begin{array}{c}
\sqrt{E+m} \sin (\theta/2) \exp(-i\phi/2)  \\
-\sqrt{E+m} \cos (\theta/2) \exp(i\phi/2)   \\
-\sqrt{E-m} \sin (\theta/2) \exp(-i\phi/2)  \\
\sqrt{E-m} \cos (\theta/2) \exp(i\phi/2)   
\end{array} 
\right) . \label{7}
\end{eqnarray}
For the state (\ref{5})--(\ref{7}), the mean values of the momentum operator, 
of the generator $\sigma^{03}/2 = i\gamma^{0} \gamma^{3}/2$ of the group 
$L^{\uparrow}_{+}$, and of the operator $W^{0} = \Sigmab {\bf P}/2$ are, respectively:
\begin{equation}
m {\bf u} = \int \bar{\psi} \left( -i\frac{\partial}{\partial{\bf r}} \right) 
\psi d{\bf r} = N \int {\bf p} \; \left[ |c_{+1}|^{2} ({\bf p}) + 
|c_{-1}|^{2} ({\bf p}) \right] \; d{\bf p}, \label{8}
\end{equation}
\begin{equation}
s^{03} = \int \bar{\psi} \left( \frac{i}{2} \gamma^{0} \gamma^{3} \right) 
\psi d{\bf r} = \frac{N}{m} \int |{\bf p}| \sin \theta \; 
{\rm Im} \left[ c_{+1}^{*} ({\bf p}) c_{-1} ({\bf p}) \right] \; d{\bf p},
\label{9}
\end{equation}
and
\begin{equation}
s^{0} = \int \bar{\psi} \left( -\frac{i}{2} \Sigmab 
\frac{\partial}{\partial{\bf r}} \right) \psi d{\bf r} = \frac{N}{2m}
\int |{\bf p}| \; \left[ |c_{+1}|^{2} ({\bf p}) - 
|c_{-1}|^{2} ({\bf p}) \right] \; d{\bf p}. \label{9a}
\end{equation}
For the sake of simplicity of the integrand expressions in (\ref{8})--(\ref{9a}) let us take the functions $c_{\pm 1}({\bf p})$ such that $c_{+1} 
({\bf p}) = ic_{-1} ({\bf p})/2 = |c(|{\bf p}|)|$. Then we conclude from 
(\ref{8}) that, for such a choice of the free state, ${\bf u} = 0$, i.e. the 
particle is at rest in the quantum mechanical sense. At the same time it 
follows from (\ref{9}) and (\ref{9a}) that neither the Frenkel's rule (\ref{2}) nor the rule  (\ref{1}) are fulfilled because both the $s^{03}$ component of the spin tensor and the $s^{0}$ component of the 4-vector are not zero.

Another example of a state for which the relation (\ref{2}) is violated 
concerns to a Dirac particle interacting with an external field. Recall the 
well-known solution of a problem of motion of an electron in a constant 
homogeneous magnetic field, whose induction ${\bf B}$ is directed along the axis
$Z$ (see, e.g., [28]). For a fixed value of the momentum projection  
$p_{z}$, the electron energy levels form a discrete spectrum with 
non-degenerate ground state $E_{0}= \sqrt{m^{2}+p_{z}^{2}}$ and 
twice-degenerate excited states $E_{n}= \sqrt{m^{2}+p_{z}^{2}+2ne_{0}B}$ (where
$e_{0}=|e|$, and $n=1,2,\ldots$). Consider the electron state with the $Y$ and 
$Z$ projections of the momentum exactly equal to zero, $p_{y}=0$, $p_{z}=0$, 
and with energy $E_{1}= \sqrt{m^{2}+2e_{0}B}$, which has the form 
\begin{equation}
\Psi = N_{1}^{1/2} \exp(-iE_{1}t) \left[ a_{+1} \Psi_{+1} + 
a_{-1} \Psi_{-1} \right], \label{10}
\end{equation}
where
\begin{eqnarray}
\Psi_{+1} = \exp(-\xi^{2}/2) \left(
\begin{array}{c}
(E_{1}+m) \\
0 \\
0 \\
2i\sqrt{e_{0}B} \xi
\end{array} 
\right) , \label{11}
\\
\Psi_{-1} = \exp(-\xi^{2}/2) \left(
\begin{array}{c}
0 \\
(E_{1}+m)\xi \\
-i\sqrt{e_{0}B}  \\
0 \\
\end{array} 
\right) , \label{12}
\end{eqnarray}
$\xi = \sqrt{e_{0}B} x$, $N_{1}$ is the normalization coefficient. We take the
constants $a_{+1}$ and $a_{-1}$ such that ${\rm Re} (a_{+1}a_{-1}^{*}) \neq 0$
and ${\rm Im} (a_{+1}a_{-1}^{*}) \neq 0$. For the state (\ref{10})--(\ref{12}), the mean values of the momentum projection onto the axis $X$, of the generator 
$\sigma^{03}/2$ of the group $L^{\uparrow}_{+}$, and of the operator $W^{0} = \Sigmab {\bf P}/2$ are, respectively:
\begin{equation}
mu_{x} = \int\limits_{-\infty}^{+\infty} \bar{\psi} \left( -i \frac{d{}}{d\xi}
\right) \psi \; d\xi = 0, \label{13}
\end{equation}
\begin{equation}
s^{03} = (e_{0}B)^{-1/2} \int\limits_{-\infty}^{+\infty} \bar{\psi} \left( 
\frac{i}{2} \gamma^{0} \gamma^{3} \right) \psi  \; d\xi
= 2\sqrt{\pi}(E_{1}+m) N_{1} {\rm Re} (a_{+1} a_{-1}^{*}) \neq 0, \label{14}
\end{equation}
and
\begin{equation}
s^{0} = \int\limits_{-\infty}^{+\infty} \bar{\psi} \left( 
-\frac{i}{2m} \Sigma_{1} \frac{d{}}{d\xi} \right) \psi  \; d\xi
= \sqrt{\pi} (E_{1}+m) N_{1} {\rm Im} (a_{+1} a_{-1}^{*}) \neq 0. \label{14a}
\end{equation}
It follows from here that, for the chosen state of an electron in the magnetic
field, the relations (\ref{2}) and (\ref{1}) are not fulfilled.

Thus, if a quantum mechanical state possesses a definite value of the 
4-momentum, then the spin tensor $s^{\mu\nu}$ and the spin 4-vector 
$s^{\alpha}$, received with averaging respectively generators $L^{\mu\nu}$ of the proper Lorentz group and operators $W^{\alpha}$ (\ref{4a}), satisfy the rules (\ref{2}) and (\ref{1}) in classical theory of spin and, consequently, give the equivalent description of the spin. The use of one or another description is determined only by reasons of simplicity in this or that situation. It is plausible that in all other states the quantum description of the spin by operators $L^{\mu\nu}$ and/or $W^{\alpha}$ breaks the rules 
(\ref{2}) and/or (\ref{1}). It means, in particular, that in the quantum theory of spin it is necessary to return to the derivation of the BMT formula, 
expressed either in terms of 4-vector $s^{\alpha}$ or tensor $s^{\mu\nu}$.

\begin{center}
{\large \bf 3. Modification of the spin rotation formula having regard to the
violation of Frenkel's rule}
\end{center}

Let us discuss now the derivation of BMT formula for the spin rotation in a 
constant homogeneous electromagnetic field $F^{\mu\nu}$. In terms of the axial 
4-vector $s^{\alpha}$, this formula looks
\begin{equation}
\frac{ds^{\alpha}}{d\tau} = \frac{ge}{2m} F^{\alpha\beta} s_{\beta} +
\frac{e}{m}\left( \frac{g}{2} - 1 \right) 
s^{\beta} F_{\beta\gamma} u^{\gamma} u^{\alpha}, \label{15}
\end{equation}
whereas in terms of the spin tensor $s^{\mu\nu}$ it has the form
\begin{eqnarray}
\frac{d s^{\mu\nu}}{d\tau} &=& \frac{ge}{2m} (s^{\mu\rho}{F^{\nu}}_{\rho} - 
s^{\nu\rho}{F^{\mu}}_{\rho}) \nonumber \\
& & + \frac{e}{m}\left( \frac{g}{2} - 1 \right)
(u^{\mu}s^{\nu\rho}-u^{\nu}s^{\mu\rho})u^{\sigma}F_{\sigma\rho}, \label{16}
\end{eqnarray}
where $\tau$ is the particle proper time, and $e$, $m$, and $g$ are the charge,
mass, and gyromagnetic ratio, respectively. Each of the formulae (\ref{15}) and
(\ref{16}) can be obtained from the other one through the relations (\ref{3}) 
and (\ref{4}) by re-expressing the quantities $s^{\alpha}$ and $s^{\mu\nu}$ in 
terms of one another (taking $a^{\mu} = 0$).

When deriving the BMT formula, it is assumed that, in the rest frame of a 
particle being in a constant homogeneous magnetic field ${\bf B}$, the 
three-dimensional vector of spin ${\bf s}$ obeys the usual equation of motion 
\begin{equation}
\frac{d{\bf s}}{d\tau} = \frac{ge}{2m} ({\bf s} \times {\bf B}). \label{17}
\end{equation}

Let us pass in Eq. (\ref{17}) from the three-dimensional vectors ${\bf s}$ and 
${\bf B}$ to their components constituting the antisymmetric tensors of the 
rotation group $s^{ij}$ and $F^{ij}$, respectively. Replace in the obtained 
equation the three-dimensional vector indices with four-dimensional Lorentz 
ones and add with arbitrary coefficients $C_{1}$ and $C_{2}$ two new terms 
vanishing in the particle rest frame where there is only a constant homogeneous 
magnetic field. Then
\begin{eqnarray}
\frac{d s^{\mu\nu}}{d\tau} &=& \frac{ge}{2m} (s^{\mu\rho}{F^{\nu}}_{\rho}-
s^{\nu\rho}{F^{\mu}}_{\rho}) + C_{1} (u^{\mu}s^{\nu\rho}-u^{\nu}s^{\mu\rho})
u^{\sigma}F_{\sigma\rho} \nonumber \\ 
& & + C_{2}(s^{\mu\rho}F^{\nu\sigma}-
s^{\nu\rho}F^{\mu\sigma})u_{\rho} u_{\sigma}. \label{18}
\end{eqnarray}

Let us look now, what occurs if to consider that the Frenkel's rule (\ref{2}) 
is true. First of all, because of it, the term with coefficient $C_{2}$ of the relation (\ref{18}) has to be omitted as equal to zero. Besides that, taking the derivative of both sides of (\ref{2}) with respect to the time $\tau$ and keeping in mind the classical equation of motion for a charged particle in an external electromagnetic field 
\begin{equation}
\frac{du^{\mu}}{d\tau} = \frac{e}{m} F^{\mu\nu}u_{\nu}, \label{19}
\end{equation}
we get 
\begin{equation}
\frac{ds^{\mu\nu}}{d\tau} u_{\nu} + \frac{e}{m} s^{\mu\nu}F_{\nu\rho}u^{\rho}
= 0. \label{20}
\end{equation}
Substituting here the expression given by the right-hand side of (\ref{18}) for 
$ds^{\mu\nu}/d\tau$, we find that
\begin{equation}
C_{1} = \frac{e}{m} \left( \frac{g}{2} - 1 \right) , \label{21}
\end{equation}
i.e. the formula (\ref{16}) generates.

The derivation of the formula (\ref{15}) for a 4-vector $s^{\alpha}$, being 
founded on the rule (\ref{1}), is similar to what was written for the tensor 
$s^{\mu\nu}$.

Since, as it has been shown above, in the quantum theory the situations are 
frequent when Frenkel's rule (\ref{2}) is not fulfilled, we believe it to be 
highly probable that this rule is violated at passing of a particle through an 
external magnetic field. This calls in question the lawfulness of the 
description of the spin rotation of a relativistic proton with the formula 
(\ref{16}) (or (\ref{15})). At a logic level, it is necessary to replace this 
formula by the modified equation (\ref{18}) with unknown coefficients $C_{1}$ 
and $C_{2}$, which are, most likely, some functions of the invariants made up 
of tensors $s^{\mu\nu}$, $F^{\mu\nu}$, and 4-vector $u^{\mu}$, and which can 
appear essentially different for the charged leptons and the baryons. 
Expressing the Eq. (\ref{18}) in terms of an axial 4-vector at violation of the
Frenkel's rule (\ref{2}) is inadmissible.

\begin{center}
{\large \bf 4. On the experimental tests of Bargmann--Michel--Telegdi formula}
\end{center}

In the absence of suitable exact solutions of the quantum theory, no 
opportunity is seen to say something certain about the quantities $C_{1}$ and 
$C_{2}$ in Eq. (\ref{18}). In such a situation, an experimental research of 
whether the BMT formula (\ref{16}) (or (\ref{15})) is a good or bad 
approximation for relativistic particles of this or that sort, is very 
important. It could be realized by the means of additional procedures in 
polarization experiments on elastic electron-proton scattering.

The main task of these experiments is in extracting the ratio of the transverse 
$P_{t}$ (in the plane of all the particle momenta) and longitudinal $P_{l}$ 
components of the recoil proton polarization vector (which is twice the rest 
spin vector). The secondary scattering on a carbon target is only sensitive to 
the transverse component of the incident particle polarization. Therefore, the
protons are passed through a magnetic dipole before getting to a carbon target. 
As a result of the spin rotation in magnetic field, the initial longitudinal
component of the recoil proton polarization vector contributes to the values of
the transverse components of polarization of the proton on its output from a 
dipole.

At the practical use of the formula (\ref{16}) (or (\ref{15})), it is accepted 
to take the induction ${\bf B}$ of a magnetic field, the particle velocity 
${\bf v}$ and the time $t$ in the laboratory frame and, using of Eq. (\ref{2}) 
(or (\ref{1})), to express the components of the spin tensor (or 4-vector) 
through its space-space part ${\bf s}$ in the particle instant rest frame. As a
result, the evolution of the rest spin of a particle is described by the 
following equation (see, e.g., [29])
\begin{equation}
\frac{d{\bf s}}{dt} = \frac{e}{m\gamma} {\bf s} \times \left[ \frac{g}{2} 
{\bf B}+\left( \frac{g}{2} -1 \right) (\gamma - 1) {\bf B}^{\perp} \right],
\label{22}
\end{equation}
where $\gamma = (1-v^{2})^{-1/2}$, and ${\bf B}^{\perp}$ is the induction 
component perpendicular to the velocity at the given instant. 

In case of violation of the Frenkel's rule  for a relativistic particle in a 
magnetic field, the formula (\ref{16}) should be replaced with the formula 
(\ref{18}), in which, generally speaking, $C_{1} \neq (e/m)(g/2 - 1)$ and 
$C_{2} \neq 0$. But introducing now the description of spin tensor in the 
particle instant rest frame is inexpedient, because it will be not more simple,
than in the laboratory frame. Both in that, and in other frame of reference, 
generally speaking, all or some of the components $s^{0i}$ ($i = 1, 2, 3$) of 
the spin tensor are nonzero. Thus, besides the uncertainty in the coefficients 
$C _ {1}$ and $C _ {2} $, the formula (\ref{18}) bears other complication, 
namely, it is not reducible to a system of three first order differential 
equations, which would be in some sense similar to the system (equation) 
(\ref{22}).

Our proposition about checking the standard spin rotation formula, expressed 
now by Eq. (\ref{22}), consists in obtaining final results of polarization 
experiments on elastic electron-proton scattering for various vectors of the 
induction ${\bf B}$ of a magnetic field.

In Ref. [4] describing the polarization experiment at Jefferson Laboratory, a sketch of the experimental set-up is presented showing the mutual arrangement of the velocity vector ${\bf v}$, the transverse component of the polarization ${\bf P}_{t}$ and the magnetic field induction ${\bf B}$ when the recoil proton enters the magnetic dipole: 
${\bf v} \perp {\bf B}$, ${\bf P}_{t} \| {\bf B}$. It would be expedient to have a set of magnetic dipoles with different directions of the induction vector ${\bf B}$ with respect to ${\bf P}_{t}$ and to ${\bf v}$ and to change the strength of the induction in a wide enough interval. If the BMT formula is true for relativistic protons, then, at the given proton energy and the given angle of departure of the recoil proton, the extracted ratio $P_{t}/P_{l}$ will not depend on the direction and the size of the induction vector in the dipole. Otherwise, it will be possible to declare that Eq. (\ref{16}) (or (\ref{15})) is unsuitable for the description of the relativistic proton spin rotation, and that the theoretical basis of the polarization experiments being carried out does not give any opportunity to extract the ratio $P_{t}/P_{l}$ and, hence, the ratio of the proton formfactors $G_{E}/G_{M}$.
 
{\bf Acknowledgments}. The author is deeply grateful to S.P. Baranov and 
V.E. Troitsky for many stimulating discussions on the considered problematics.

\end{small}
\end{document}